\documentclass[a4paper,fleqn]{cas-dc}

\usepackage[english]{babel}
\usepackage[numbers,sort&compress]{natbib}
\usepackage{graphicx}
\usepackage{graphics}
\usepackage{braket}
\usepackage{bbold}
\usepackage{amssymb}
\usepackage{amsmath}
\usepackage{nicefrac}
\usepackage{dcolumn}
\usepackage{bm}
\usepackage{slashed}
\usepackage{datetime}
\usepackage{mciteplus}
\usepackage{multirow}
\usepackage{bbold}
\usepackage{color}
\usepackage{xcolor}
\usepackage{float}
\usepackage[utf8]{inputenc}
\usepackage[normalem]{ulem}

\def\tsc#1{\csdef{#1}{\textsc{\lowercase{#1}}\xspace}}
\tsc{WGM}
\tsc{QE}

\newcommand{\drv}{{\rm d}}

\newcommand{\MSb}{\overline{\rm MS}}

\newcommand{\CmHENLO}{C_n^{{\rm HE}\text{-}{\rm NLO}}}
\newcommand{\DY}{\Delta Y}
\newcommand{\JPsi}{J/\psi}

\newcommand{\XQq}{ X_{Qq\bar{Q}\bar{q}}}
\newcommand{\Xcu}{ X_{cu\bar{c}\bar{u}}}
\newcommand{\Xcs}{ X_{cs\bar{c}\bar{s}}}
\newcommand{\Xbu}{ X_{bu\bar{b}\bar{u}}}
\newcommand{\Xbs}{ X_{bs\bar{b}\bar{s}}}

\newcommand{\tcite}[1]{~\cite{#1}}
\newcommand{\tref}[1]{~\ref{#1}}
\newcommand{\eref}[1]{~\eqref{#1}}

\begin{document}
\let\WriteBookmarks\relax
\def\floatpagepagefraction{1}
\def\textpagefraction{.001}

\shorttitle{A high-energy QCD portal to exotic matter: Heavy-light tetraquarks at the HL-LHC}    

\shortauthors{Celiberto, Francesco Giovanni}
\shortauthors{Papa, Alessandro} 

\title []{\Huge A high-energy QCD portal to exotic matter: \\ Heavy-light tetraquarks at the HL-LHC}  

\author[1]{Francesco Giovanni Celiberto}[orcid=0000-0003-3299-2203]
\cormark[1]
\author[2,3]{Alessandro Papa}[orcid=0000-0001-8984-3036]


\ead{francesco.celiberto@uah.es}
\ead{alessandro.papa@fis.unical.it}


\affiliation[1]{organization={Universidad de Alcal\'a (UAH), Departamento de F\'isica y Matem\'aticas},
            addressline={Campus Universitario}, 
            city={Alcal\'a de Henares},
            postcode={E-28805}, 
            state={Madrid},
            country={Spain}}

\affiliation[2]{organization={Dipartimento di Fisica, Universit\`a della Calabria},
            addressline={Ponte Pietro Bucci, Cubo 31C}, 
            city={Arcavacata di Rende},
            postcode={I-87036}, 
            state={Cosenza},
            country={Italy}}

\affiliation[3]{organization={INFN, Gruppo Collegato di Cosenza},
            addressline={Ponte Pietro Bucci, Cubo 31C}, 
            city={Arcavacata di Rende},
            postcode={I-87036}, 
            state={Cosenza},
            country={Italy}}

\cortext[1]{Corresponding author.}



\begin{abstract}
By taking advantage of the natural stability of the high-energy resummation, recently discovered in the context of heavy-flavor studies, we investigate the inclusive hadroproduction of a neutral heavy-light, hidden-flavored tetraquark ($\Xcu$ or $\Xbs$ state), in association with a heavy (single $c$- or $b$-flavored) hadron or a light jet at the (HL-)LHC.
We make use of the {\tt JETHAD} multi-modular working package to provide predictions for rapidity, azimuthal-angle and transverse-momentum distributions calculated
\emph{via} the hybrid high-energy and collinear factorization, where the  Balitsky--Fadin--Kuraev--Lipatov resummation of energy logarithms is supplemented by collinear parton densities and fragmentation functions.
We rely upon the single-parton fragmentation mechanism, valid in the large transverse-momentum regime, to describe the tetraquark production. Our study represents a first attempt at bridging the gap between all-order calculations of high-energy QCD and the exotics.
\end{abstract}




\maketitle

\section{Hors d'{\oe}uvre}
\label{sec:introduction}

Can the production of a tetraquark at hadron colliders be described within high-energy Quantum Chromodynamics (QCD)?
While unveiling the core dynamics leading to exotic matter formation still lies over the horizon of frontier researches, recent advancements in all-order perturbative techniques and QCD factorization(s) may open novel and unexpected perspectives.

Though mesons and baryons represent the simplest valence-quark combinations which can form colorless particles, and we call them \emph{ordinary} hadrons, QCD color neutrality does not prohibit the formation of bound states with different valence-parton configurations.
Since quantum numbers of these hadrons generally cannot be reproduced by ordinary particles, we call them \emph{exotics} (see~\cite{Esposito:2016noz,Olsen:2017bmm,Brambilla:2019esw,Karliner:2017qhf,Lebed:2016hpi} for a review).
The enigmatic description of their inner structure has been matter of intense investigation by exotic spectroscopy (see, \emph{e.g.}~\cite{Ferretti:2018ojb,Ferretti:2020ewe,Guo:2009id} and references therein).
Two main kinds of structures have been proposed: $(i)$ lowest Fock states with active gluons, like quark-gluon \emph{hybrids} and \emph{glueballs}, or $(ii)$ \emph{multi-quarks}, like tetraquarks and pentaquarks.

The discovery of the first exotic, the $X(3872)$, happened in 2003 at Belle~\cite{Belle:2003nnu}, and it was then confirmed by several other experiments.
The $X(3872)$ is a hidden-flavored particle, namely composed by pair(s) of heavy-flavored quarks~\cite{Chen:2016qju,Liu:2019zoy,Esposito:2020ywk}.\footnote{
The first open-charm exotic, named $X(2900)$, was detected in 2021 at LHCb~\cite{LHCb:2020bls}.}
Its discovery marked the turn of the so-called second quarkonium revolution (the first revolution coincided with $\JPsi$ observation in 1974).
Although $X(3872)$ possesses non-exotic quantum numbers, its decays violate isospin.
Therefore, other dynamical mechanisms beyond the pure quarkonium scenario and closer to a tetraquark-like description  have been proposed so far.
A tetraquark state may be: a loosely-bound meson molecule~\cite{Guo:2017jvc}, a compact double di-quark system~\cite{Maiani:2004vq}, or a hadroquarkonium made by a quarkonium core plus an orbiting light meson~\cite{Dubynskiy:2008mq}.
The quite large $X(3872)$ production rates measured by LHC collaborations at large transverse momentum~\cite{CMS:2013fpt,ATLAS:2016kwu,LHCb:2021ten}  could set constrains on the validity of production mechanisms, in particular on the molecular vision~\cite{Bignamini:2009sk,Artoisenet:2009wk,Guerrieri:2014gfa,Jin:2016vjn}.
At the same time, they could favor other mechanisms natively embodied by high-energy QCD, such as the \emph{fragmentation} of a single parton into the observed particle.

More in general, emissions of hadrons with heavy quarks are useful channels to probe high-energy QCD. In this kinematic limit, logarithms of the center-of-mass energy, $\sqrt{s}$, are very large. They enter the running-coupling expansion to all orders, up to spoil the QCD perturbative convergence.
The Balitsky--Fadin--Kuraev--Lipatov (BFKL) resummation~\cite{Fadin:1975cb,Kuraev:1977fs,Balitsky:1978ic} permits to account for these logarithms to all orders. Its validity is proven up to the leading logarithmic level (LL), \emph{i.e.} the resummation of all terms proportional to $(\alpha_s \ln (s))^n$, and to the next-to-leading level (NLL), \emph{i.e.} the resummation of all contributions proportional to $\alpha_s (\alpha_s \ln (s))^n$.
BFKL distributions for hadronic processes take the form of a convolution between a process-universal Green's function, which resums towers of high-energy logarithms~\cite{Fadin:1998py,Ciafaloni:1998gs,Fadin:2004zq,Fadin:2023roz} and is known at next-to-leading order (NLO), and two process-dependent impact factors. Only a few of them are known at NLO. They embody collinear ingredients such as parton densities (PDFs) and fragmentation functions (FFs). Their presence makes our formalism be a \emph{hybrid} high-energy and collinear factorization.

BFKL has been tested \emph{via} a series of phenomenological studies: Mueller--Navelet productions~\cite{Mueller:1986ey,Colferai:2010wu,Ducloue:2013bva,Caporale:2014gpa,Celiberto:2015yba,Celiberto:2022gji,deLeon:2021ecb}, light di-hadron tags~\cite{Bolognino:2018oth,Celiberto:2022kxx}, Higgs plus jet~\cite{Celiberto:2020tmb}, forward Drell--Yan~\cite{Celiberto:2018muu,Golec-Biernat:2018kem}, and heavy-hadron detections~\cite{Boussarie:2017oae,Celiberto:2017nyx,Bolognino:2019ouc,Bolognino:2021mrc,Celiberto:2021dzy,Celiberto:2021fdp,Celiberto:2022dyf,Celiberto:2022keu}.
Probing BFKL through heavy flavor permits, as a main advantage, to overcome the well-known instabilities rising when light-flavored jets or hadrons are detected.
Those issues are partially related to NLL corrections both
to Green’s function and impact factors, which
are of same size and opposite sign than LL terms.
They are also generated by large double \emph{threshold} logarithms, emerging at large rapidity intervals, not accounted for by our formalism.
As a result, the high-energy series becomes unstable under renormalization and factorization scale variations around \emph{natural} energies given by kinematic, this preventing any chance of making precision studies~\cite{Ducloue:2013bva,Caporale:2014gpa,Celiberto:2020wpk}.

Recent works on single heavy-flavored particles, such as $\Lambda_c$ baryons~\cite{Celiberto:2021dzy} or $b$-hadrons~\cite{Celiberto:2021fdp}, have shown how variable-flavor-number-scheme (VFNS)~\cite{Buza:1996wv} collinear FFs describing parton hadronization to heavy-quark bound states at large transverse momentum brings to a \emph{natural stabilization} of the resummation~\cite{Celiberto:2022grc}.
Looking for clear signals of a systematic stabilizing pattern of high-energy dynamics \emph{via} the production of heavy hadrons, the fragmentation approximation was applied also to vector quarkonia~\cite{Celiberto:2022dyf} and charmed $B$ mesons~\cite{Celiberto:2022dyf}.
There, VFNS, DGLAP-evolving FFs were built on the basis of non-relativistic QCD (NRQCD)~\cite{Caswell:1985ui,Bodwin:1994jh} initial-scale inputs at NLO~\cite{Zheng:2019dfk,Zheng:2019gnb}.
As a result, natural stability passed the NRQCD stress test, with a corroborating evidence from both quarkonium and $B_c$ channels.

In this work we make use of the natural stability to enlarge our horizons toward a prime description of neutral heavy-light, hidden-flavored tetraquarks within high-energy QCD at NLL/NLO.
By relying upon the single-parton fragmentation mechanism, we will first build a novel set of tetraquark VFNS FFs which evolve with DGLAP.
Then we will plug them into our hybrid factorization.\footnote{For a complementary high-energy approach to the single hadroproduction of (fully) charmed tetraquarks, see~\cite{Maciula:2020wri,Cisek:2022uqx}.}
A very clean channel to study tetraquark production would be the inclusive single-forward emission. To perform such a study via BFKL, we should rely upon the small-$x$ unintegrated gluon density (UGD) in the proton. However, our current knowledge of the UGD is very qualitative and comes mostly from model-dependent studies (see, \emph{e.g.},~\cite{Bolognino:2018rhb,Bolognino:2021niq}).
Conversely, the main advantage of producing a jet in association with the tetraquark is that, when the two objects are emitted in a forward/backward configuration, secondary emissions between them are expected to be strongly ordered in rapidity. This permits us to easily access the BFKL dynamics in a moderate-$x$ regime, where collinear PDFs, much better known than the UGD, can be used. 
Therefore, here we focus on tetraquark-plus-jet observables and we postpone single forward emissions to future studies.

We propose this study without pretension of catching the core features of tetraquark production by the hands of BFKL, but rather providing a novel and complementary channel to access the physics of exotics thanks to high-energy techniques.
Our analysis will serve as a portal for future phenomenological efforts to shed light on tetraquark formation at the LHC and its high-luminosity (HL) upgrade.

\section{Hybrid high-energy and collinear factorization}
\label{sec:theory}

\begin{figure*}[tb]
\centering

\includegraphics[scale=0.45,clip]{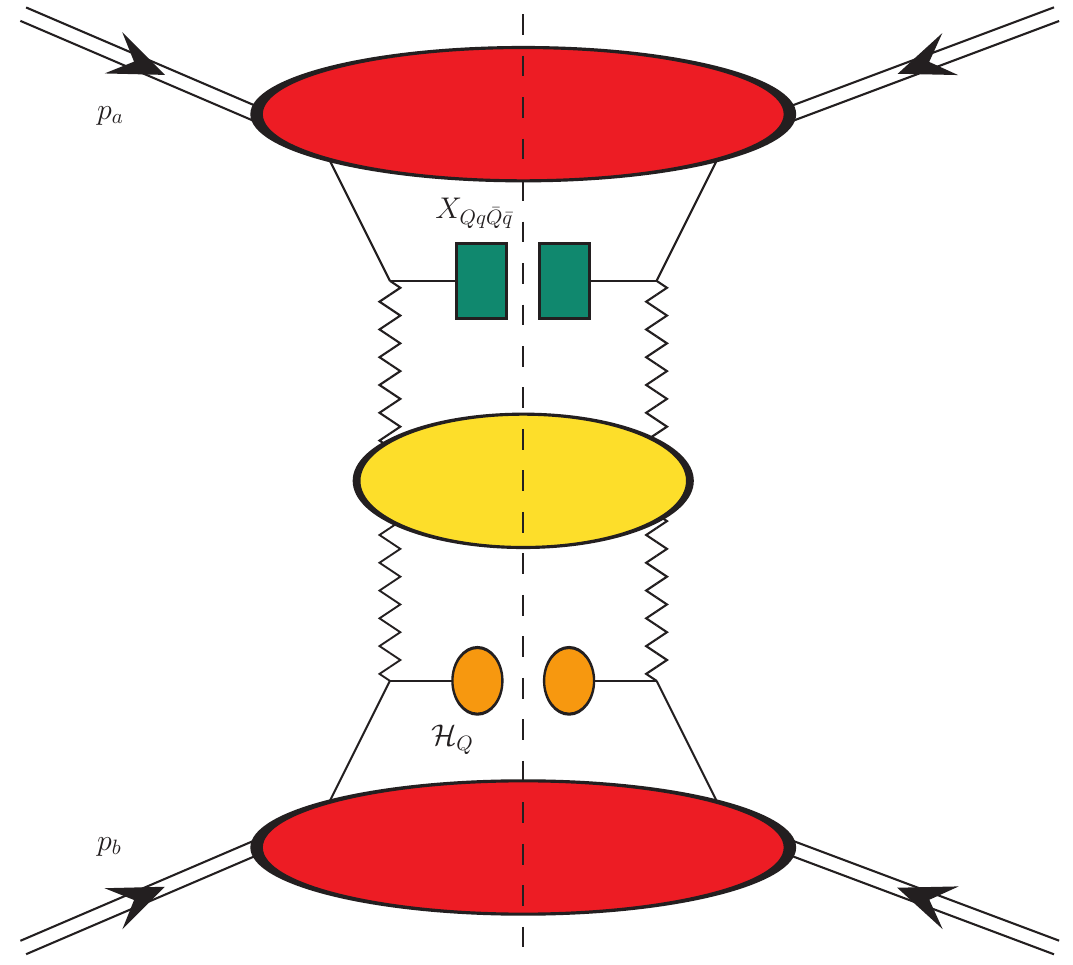}
\hspace{0.50cm}
\includegraphics[scale=0.45,clip]{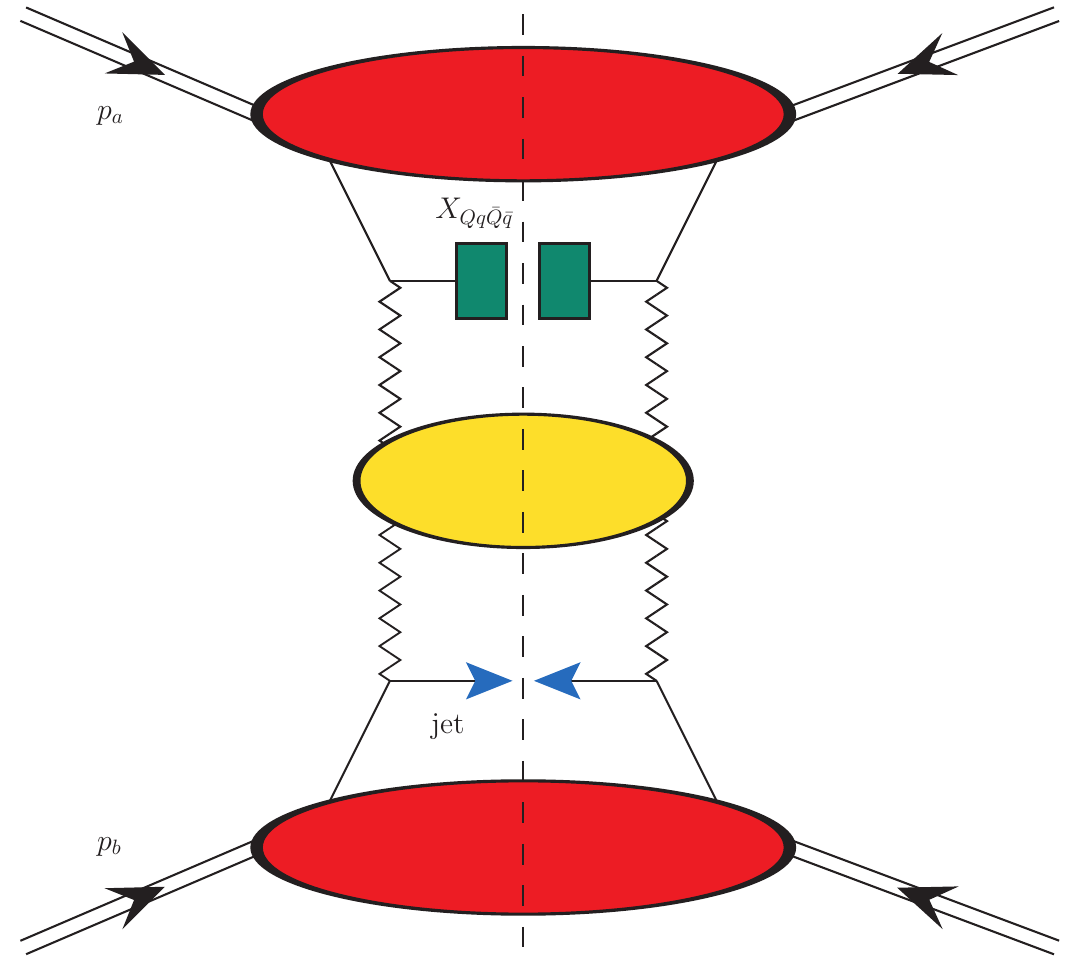}

\caption{Diagrammatic view of the hybrid high-energy and collinear factorization for tetraquark-plus-hadron (left) and tetraquark-plus-jet (right) hadroproductions. Red blobs depict incoming-proton collinear PDFs. Green rectangles stands for tetraquark collinear FFs. Orange ovals portray single-charmed hadrons. Blue arrows denote light jets. The BFKL Green's function, given by the big yellow blob, is connected to the two impact factors by Reggeon lines. Diagrams were made through {\tt JaxoDraw 2.0}\tcite{Binosi:2008ig}.}
\label{fig:process}
\end{figure*}

The reaction under investigation is (see Fig.\tref{fig:process})
\begin{eqnarray}
\label{process}
 {\rm p}(p_a) + {\rm p}(p_b) \to \XQq(\kappa_1, y_1) + {\cal X} + {\cal O}(\kappa_2, y_2) \;,
\end{eqnarray}
where an outgoing heavy-light tetraquark, $\Xcu$ or $\Xbs$ is accompanied by a single $c$- or $b$-flavored hadron, or a light jet, ${\cal O} = \{ {\cal H}_c, {\cal H}_b, {\rm jet} \}$, and together with an undetected system, ${\cal X}$. The final-state objects feature transverse momenta, $|\bm{\kappa}_{1,2}| \gg \Lambda_{\rm QCD}$, and they are separated by a large distance in rapidity, $ \DY = y_1 - y_2$. 
A ${\cal H}_c$ particle is an inclusive state consisting in the sum of fragmentation channels to single-charmed $D^\pm$, $D^0$ and $D^{*\pm}$~mesons, as well as $\Lambda_c^\pm$ baryons.
Conversely, a ${\cal H}_b$ hadron is build as the sum of non-charmed $B$~mesons and $\Lambda_b^0$ baryons, see~\cite{Celiberto:2021fdp}.
We use a Sudakov decomposition of the $\kappa_{1,2}$ four-momenta on the basis of the colliding-proton momenta, $p_{a,b}$, to get
\begin{eqnarray}
\label{sudakov}
\kappa_{1,2} = x_{1,2} p_{a,b} + \frac{\bm{\kappa}_{1,2}^2}{x_{1,2} s}p_{b,a} + \kappa_{1,2\perp} \ , \quad
\kappa_{1,2\perp}^2 = - \bm{\kappa}_{1,2}^2 \;,
\end{eqnarray}
In the center-of-mass frame one has
\begin{eqnarray*}
\label{rapidities}
y_{1,2} = \pm \ln  \frac{x_{1,2} \sqrt{s}}{|\bm{\kappa}_{1,2}| \;,
}
\end{eqnarray*}
with $x_{1,2}$ the final-state longitudinal momentum fractions.

\subsection{NLL-resummed cross section}
\label{sec:cross_section}

We write the $\DY$- and $\varphi$-differential cross section, with $\varphi = \varphi_1 - \varphi_2 - \pi$ and $\varphi_{1,2}$ the azimuthal angles of the two identified objects in the final state, as a Fourier sum of azimuthal coefficients, $C_{n \ge 0}$
\begin{eqnarray}
 \label{dsigma_Fourier}
 \frac{\drv \sigma}{\drv \DY \, \drv \varphi \, \drv |\bm{\kappa}_1| \, \drv |\bm{\kappa}_2|} 
 =
 \frac{1}{2\pi} \left[C_0 + 2 \sum_{n=1}^\infty \cos (n \varphi)\,
 C_n \right]\;,
\end{eqnarray}
Working in the hybrid high-energy and collinear factorization and in the $\MSb$ renormalization scheme, we draw a master expression for the $C_n$ coefficients~\cite{Caporale:2012ih}. It is valid within the NLO perturbative expansion and encodes the NLL resummation of high-energy logarithms:
\[
 \hspace{-0.65cm}
 C_n^{\rm NLL} = 
 \int_{y_1^{\rm min}}^{y_1^{\rm max}} 
 \drv y_1
 \int_{y_2^{\rm min}}^{y_2^{\rm max}} 
 \drv y_2
 \; \delta(\DY - y_1 + y_2)
 \; \frac{e^{\DY}}{s}
\]
\[
 \hspace{-0.65cm}
 \times \;
 \int_{-\infty}^{+\infty} \hspace{-0.20cm} \drv \nu \, e^{\bar \alpha_s \DY \chi^{\rm NLO}(n,\nu)}
 \, \alpha_s^2(\mu_R)
 \bigg[ 
 \bar \alpha_s^2 \frac{\beta_0 \DY}{4 N_c}\chi(n,\nu)f(\nu)
\]
\begin{eqnarray}
\label{Cn_NLL_MSb}
 \hspace{-0.65cm}
 + \,\,
 \Phi_1^{\rm NLO}(n,\nu,|\bm{\kappa}_1|, x_1)[\Phi_2^{\rm NLO}(n,\nu,|\bm{\kappa}_2|,x_2)]^*
 \bigg] \;,
\end{eqnarray}
where $\bar \alpha_s(\mu_R) = \alpha_s(\mu_R) N_c/\pi$, $N_c$ is the color number, and $\beta_0 = 11N_c/3 - 2 n_f/3$.
The BFKL kernel at the exponent in~\eqref{Cn_NLL_MSb} resums the NLL energy logarithms
\begin{eqnarray}
 \label{chi}
 \chi^{\rm NLO}(n,\nu) = \chi(n,\nu) + \bar\alpha_s \hat \chi(n,\nu) \;,
\end{eqnarray}
with $\chi(n,\nu)$ the leading-order (LO) BFKL eigenvalues
\begin{eqnarray}
\chi\left(n,\nu\right) = -2\left\{\gamma_{\rm E}+{\rm Re} \left[\psi\left( (n + 1)/2 + i \nu \right)\right] \right\} \;,
\label{chi_LO}
\end{eqnarray}
then $\psi(z) = \Gamma^\prime(z)/\Gamma(z)$ and $\gamma_{\rm E}$ the Euler-Mascheroni constant.
The $\hat\chi(n,\nu)$ function is the NLO kernel correction
\begin{align}
\label{chi_NLO}
\hat \chi\left(n,\nu\right) &= \bar\chi(n,\nu)+\frac{\beta_0}{8 N_c}\chi(n,\nu)
\\ \nonumber &\times \,
\left\{-\chi(n,\nu)+10/3+2\ln\left(\mu_R^2/\hat{\mu}^2\right)\right\} \;,
\end{align}
with the characteristic function  $\bar\chi(n,\nu)$ calculated in~\cite{Kotikov:2000pm} and $\hat{\mu} = \sqrt{|\bm{\kappa}_1| |\bm{\kappa}_2|}$. The two functions
\begin{eqnarray}
\label{IFs}
\Phi_{1,2}^{\rm NLO}(n,\nu,|\bm{\kappa}|,x) =
\Phi_{1,2} +
\alpha_s(\mu_R) \, \hat \Phi_{1,2}
\end{eqnarray}
are the NLO impact factors.
Emissions of our tetraquarks as well as of the ${\cal H}_{Q} \equiv {\cal H}_{c,b}$ particles are described by the NLO forward-hadron impact factor~\cite{Ivanov:2012iv}. Although designed for the study of light hadrons, its use is also valid in our VFNS approach, provided that transverse-momenta ranges are much larger than heavy-quark DGLAP-evolution thresholds.
The LO hadron impact factor encodes the collinear convolution
\begin{eqnarray}
\nonumber
\Phi_H(n,\nu,|\bm{\kappa}|,x) = \rho_c \, |\bm{\kappa}|^{2i\nu-1}\int_{x}^1 \drv \zeta / \zeta \; \hat{x}^{1-2i\nu} 
\nonumber \\
\label{LOHIF}
 \times \, \Big[\tau_c f_g(\zeta)D_g^H\left(\hat{x}\right)
 +\sum_{i=q,\bar q}f_i(\zeta)D_i^H\left(\hat{x}\right)\Big] \;,
\end{eqnarray}
with $\hat{x} = x/\zeta$, $\rho_c = 2 \sqrt{C_F/C_A}$, and $\tau_c = C_A/C_F$, where $C_F = (N_c^2-1)/(2N_c)$ and $C_A = N_c$. Then, $f_i\left(x, \mu_F \right)$ is the PDF for the parton $i$ extracted from the parent proton, whereas $D_i^H\left(x/\beta, \mu_F \right)$ is the FF for the parton $i$ fragmenting to the tagged hadron, $H$.
The NLO correction can be found in~\cite{Ivanov:2012iv}.
The LO forward-jet impact factor reads
\begin{eqnarray}
 \label{LOJIF}
 \hspace{-0.09cm}
 \Phi_J(n,\nu,|\bm{\kappa}|,x) = \rho_c
 |\bm{\kappa}|^{2i\nu-1}\,\hspace{-0.05cm}\Big(\tau_c f_g(x)
 +\hspace{-0.15cm}\sum_{j=q,\bar q}\hspace{-0.10cm}f_j(x)\Big) \;,
\end{eqnarray}
while its NLO correction is obtained by combining Eq.~(36) of~\cite{Caporale:2012ih} with Eqs.~(4.19)-(4.20) of~\cite{Colferai:2015zfa}.
It relies upon a small-cone selection functions~\cite{Ivanov:2012ms} with the jet-cone radius fixed at $r_J = 0.5$, as adopted in recent analyses at CMS~\cite{Khachatryan:2016udy}.
The expression for the $f(\nu)$ function in~\eqref{Cn_NLL_MSb} is
\begin{eqnarray}
 f(\nu) = \left[ \frac{i}{2} \frac{\drv}{\drv \nu} \ln\frac{\Phi_1}{\Phi_2^*} + 2 \ln \hat{\mu} \right] \;.
\label{fnu}
\end{eqnarray}
The cross product between the two NLO impact-factor parts in Eq.\eref{Cn_NLL_MSb} represents a next-to-NLO correction which will not be taken into account in the final implementation of our NLL/NLO master formula.

A comprehensive high-energy versus fixed-order analysis builds on confronting NLL-resummed predictions with pure fixed-order calculations.
However, according to our knowledge, a numerical tool to compute NLO observables sensitive to two-particle hadroproductions is not yet available.
To get a reference fixed-order calculation, we truncate the expansion of $C_n$ coefficients in~(\ref{Cn_NLL_MSb}) up to ${\cal O}(\alpha_s^3)$. Thus, we come out with an effective high-energy fixed-order (HE-NLO) formula, suited to phenomenology.
It collects the leading-power asymptotic signal present in a pure NLO calculation, while factors suppressed by inverse powers of the partonic center-of-mass energy are discarded.
The $\MSb$ expressions for HE-NLO azimuthal coefficients reads
\begin{eqnarray}
\label{Cn_HENLO_MSbar}
 \hspace{-0.65cm}
 \CmHENLO = 
 \int_{y_1^{\rm min}}^{y_1^{\rm max}} 
 \!\!\! \drv y_1
 \int_{y_2^{\rm min}}^{y_2^{\rm max}} 
 \!\!\! \drv y_2
 \; \delta(\DY - y_1 + y_2)
 \; \frac{e^{\DY}}{s}
\end{eqnarray}
\[
 \hspace{-0.65cm}
 \times \;
 \,
 \int_{-\infty}^{+\infty} \hspace{-0.20cm} \drv \nu \,
 \alpha_s^2(\mu_R) \, \Phi_1(n,\nu,|\bm{\kappa}_1|, x_1)[\Phi_2(n,\nu,|\bm{\kappa}_2|,x_2)]^*
\]
\[
 \hspace{-0.65cm}
 \times \; \big\{ 
 1 + \bar \alpha_s(\mu_R) \DY \chi(n,\nu) 
\]
\[
 \hspace{-0.65cm}
 + \; \alpha_s(\mu_R) \big[ \Phi_1^{\rm NLO}/\Phi_1 + \big[ \Phi_2^{\rm NLO}/\Phi_2 \big]^* \big]
 \big\} \;,
\]
with the exponentiated BFKL kernel expanded up to ${\cal O}(\alpha_s)$.
The LO/LL limit is got by dropping NLO pieces in~\eqref{chi_NLO}-\eqref{IFs}.

\subsection{Collinear ingredients}
\label{ssec:PDFs_FFs}

\begin{figure*}[!t]
\centering

   \includegraphics[scale=0.51,clip]{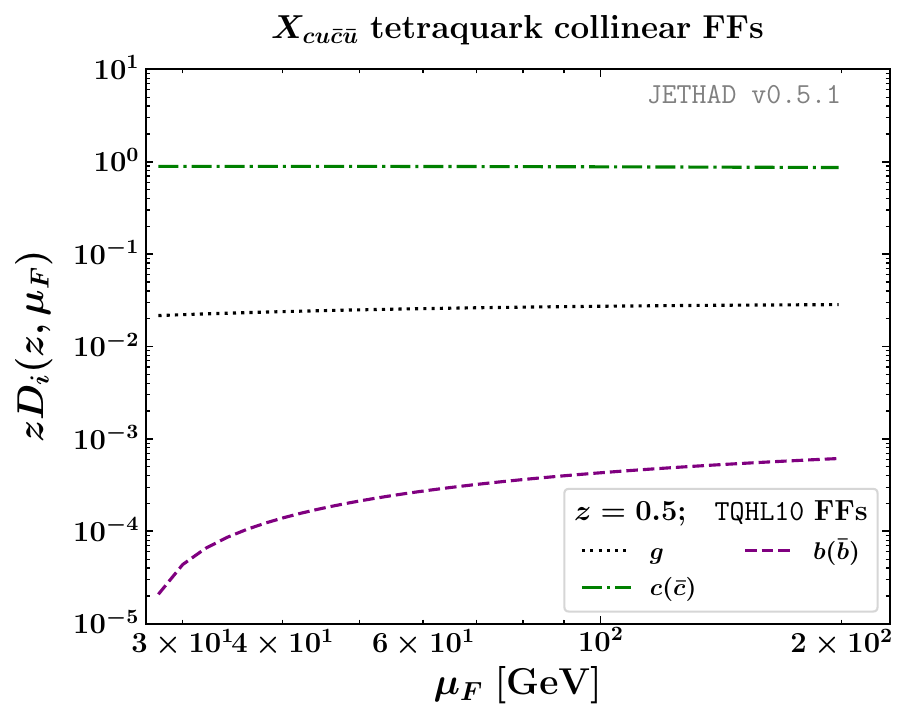}
   \hspace{0.10cm}
   \includegraphics[scale=0.51,clip]{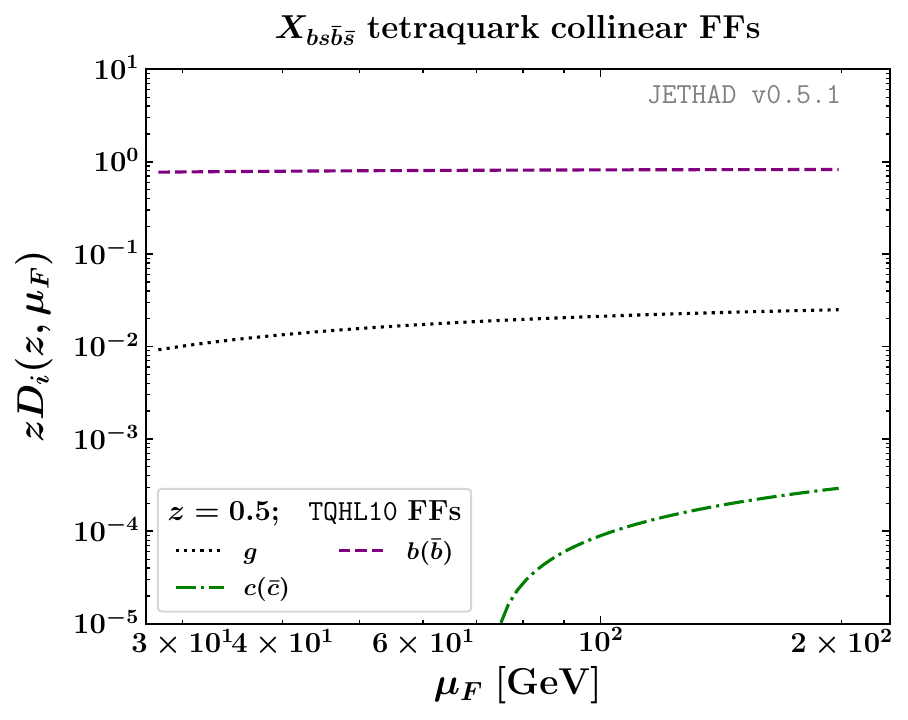}

\caption{Energy dependence of $\Xcu$ (left) and $\Xbs$ (right) {\tt TQHL1.0} NLO FFs at $z \equiv \langle z \rangle \simeq 0.45$. Light-quark species are not shown, since their contribution is negligible with respect to gluon and heavy quarks.}
\label{fig:FFs}
\end{figure*}

We fix $\mu_R$ and $\mu_F$ at the \emph{natural} energy scales of the process. We set $\mu_F = \mu_R = \mu_N = M_{1 \perp} + M_{2 \perp}$, with $M_{i \perp}^2 = M_i^2 + \bm{\kappa}_i^2$ the transverse mass of the observed $i$ particle (see, \emph{e.g.},~\cite{Alioli:2010xd,Campbell:2012am,Hamilton:2012rf}).
Tetraquark masses are set to $M_X = 2(M_Q+m_q)$, according to~\cite{Nejad:2021mmp}. Here, $M_Q$ ($m_q$) is the heavy (light) constituent quark, $Q$ ($q$).
Single heavy-flavored hadron masses are $M_{{\cal H}_c} \equiv M_{\Lambda_c} = 2.286$ GeV and $M_{{\cal H}_b} \equiv M_{\Lambda_b} = 5.62$ GeV. Since in our treatment jet mass corrections are neglected, we can safely fix $M_{2 \perp} \equiv |\bm{\kappa}_2|$ in jet emissions.
We employ the new {\tt NNPDF4.0} NLO PDF determination~\cite{NNPDF:2021uiq,NNPDF:2021njg}, obtained in a neural-network framework~\cite{Forte:2002fg}.
To depict ${\cal H}_c$ emissions, we have built a new NLO parametrization in {\tt LHAPDF} format~\cite{Buckley:2014ana}, labeled as {\tt HCFF1.0}, obtained by combining {\tt KKKS08} NLO FFs~\cite{Kniehl:2005de,Kneesch:2007ey} for $D$ mesons with {\tt KKSS19} ones~\cite{Kniehl:2020szu} for $\Lambda_c$ baryons.
${\cal H}_b$ hadrons are described by the {\tt KKSS07} NLO set~\cite{Kniehl:2011bk,Kramer:2018vde}.
Our starting point to construct a DGLAP-evolving tetraquark FF set is the calculation of the \emph{direct} $(Q \to \XQq)$ $S$-wave collinear function done by Moosavi Nejad and Amiri~\cite{Nejad:2021mmp}. It is based on the spin-dependent (and thus, transverse-momentum-dependent) Suzuki model~\cite{Suzuki:1985up} (see also~\cite{Suzuki:1977km,Amiri:1986zv}). Then, to obtain the collinear limit, the relative motion of constituent quark inside the tetraquark is neglected~\cite{Lepage:1980fj,Brodsky:1985cr}. The factorization treatment of the initial-scale input for tetraquark fragmentation is analogous to the one for quarkonia prescribed by NRQCD.
There, first a constituent $(Q\bar{Q})$ pair is produced perturbatively, then the quarkonium formation is described by nonperturbative long-distance matrix elements.
In our case, a $(Qq\bar{Q}q$) system is first emitted \emph{via} perturbative splittings. 
Its production amplitude is then convoluted with a bound-state wave function encoding the non-perturbative dynamics of tetraquark formation, according to the Suzuki picture.
The model in~\cite{Nejad:2021mmp} is not tuned on data, such as $X(3872)$ cross sections, but it is normalized as in~\cite{Amiri:1986zv,Suzuki:1985up}.

Starting from the input of~\cite{Nejad:2021mmp}, taken at the initial scale $\mu_0=M_X+M_Q$, and making use of the {\tt APFEL} code~\cite{Bertone:2013vaa}, we have built a first and novel 
DGLAP-evolved \emph{TetraQuarks with Heavy and Light flavors} ({\tt TQHL1.0}) FF set in {\tt LHAPDF} format, ready for phenomenology.
One might argue that our approach misses the initial-scale contribution of light partons as well as of the non-constituent heavy quark, which are generated at $\mu_F > \mu_0$ by evolution only.
However, as pointed out in~\cite{Nejad:2021mmp}, these channels are negligible at $\mu_0$. This is also true in the case of vector-quarkonium FFs, see~\cite{Celiberto:2022dyf}.

We show in Fig.\tref{fig:FFs}
the $\mu_F$-dependence of {\tt TQHL1.0} FFs
for a momentum fraction $z$ roughly corresponding to its average value, $\langle z \rangle \simeq 0.5$.
As recently shown~\cite{Celiberto:2021dzy,Celiberto:2021fdp,Celiberto:2022dyf,Celiberto:2022keu}, in the hybrid factorization the gluon-to-hadron fragmentation channel plays a critical role. Its impact on cross sections is heightened by the collinear convolution with the gluon PDF in the LO impact factor (Eq.~\eqref{LOHIF}). This holds also at NLO, where non-diagonal $(gQ)$ channels are open.
Non-decreasing with $\mu_F$ gluon FFs, as ones of Fig.~\ref{fig:FFs}, compensate the falloff with $\mu_R$ of the running coupling in the
exponentiated kernel and in impact factors. This leads to the \emph{natural stability} observed in heavy-flavor high-energy cross sections~\cite{Celiberto:2021dzy,Celiberto:2021fdp,Celiberto:2022grc}.
This remarkable property comes out as an \emph{intrinsic} feature shared by all the heavy-flavor emissions investigated so far: single $Q$-flavored hadrons~\cite{Celiberto:2021dzy,Celiberto:2021fdp,Celiberto:2022rfj,Celiberto:2022zdg}, quarkonia~\cite{Celiberto:2022dyf}, $B_c^{(*)}$ mesons~\cite{Celiberto:2022keu}, and now heavy-light tetraquarks.

\section{Phenomenology}
\label{sec:pheno}

We employed the {\tt JETHAD} multi-modular interface~\cite{Celiberto:2020wpk,Celiberto:2022rfj} for our phenomenology.
We assessed the sensitivity of our distributions on scale variations by letting $\mu_R$ and $\mu_F$ be around their natural values, up to a factor $C_\mu = \mu_{R,F}/\mu_N$ ranging from 1/2 to two. 
Shaded bands in our figures embody the comprehensive effect of scale variations and phase-space multi-dimensional integrations, this latter being uniformly kept below 1\% by {\tt JETHAD} integration algorithms.
The center-of-mass energy is $\sqrt{s} = 14$ TeV for all the observables.
Hadrons are reconstructed by the CMS barrel only, 
$|y_{1,2}| < 2.4$, while jets also by CMS endcaps~\cite{Khachatryan:2016udy}, $|y_2| < 4.7$.

\subsection{Rapidity and azimuthal distributions}
\label{ssec:DY_phi}

\begin{figure*}[!t]
\centering

   \includegraphics[scale=0.41,clip]{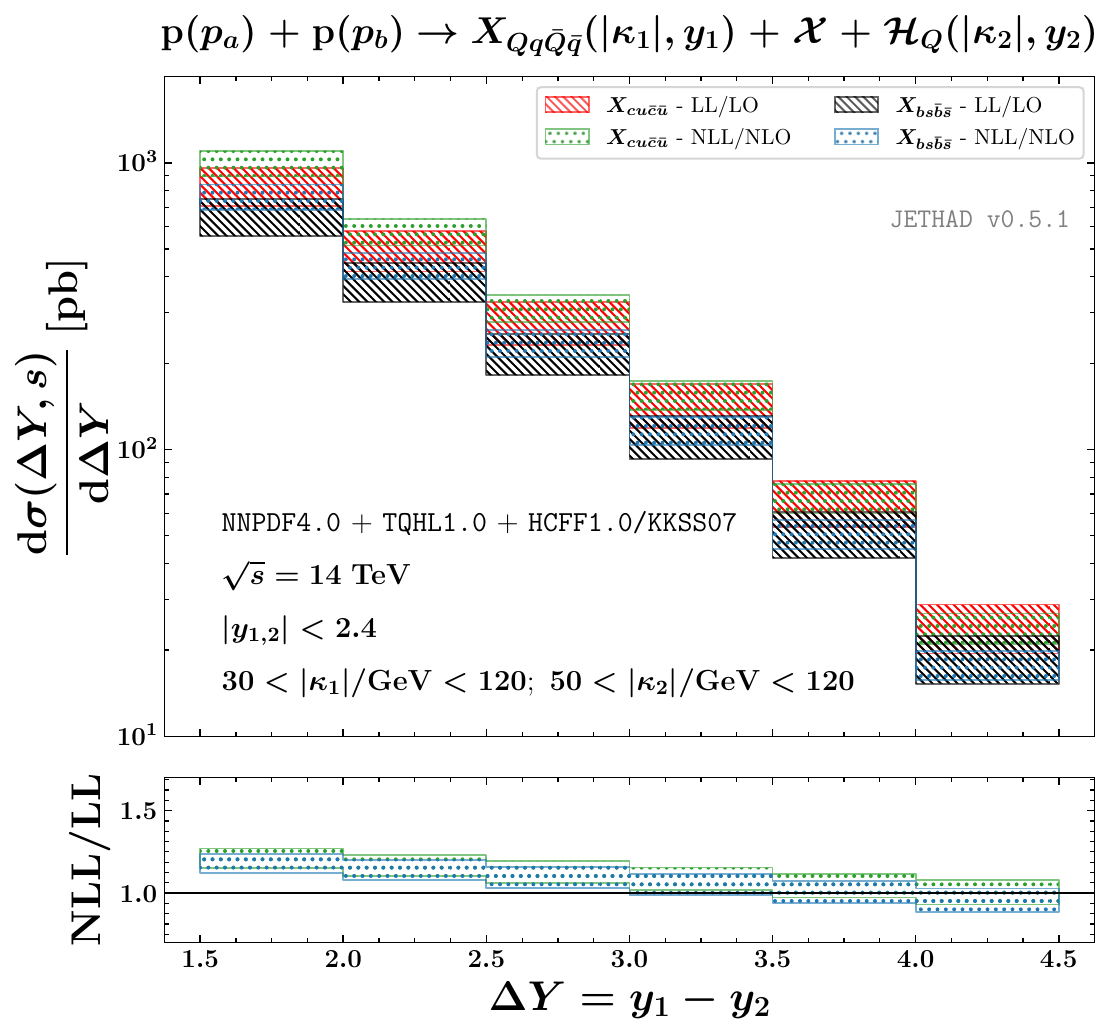}
   \hspace{0.10cm}
   \includegraphics[scale=0.41,clip]{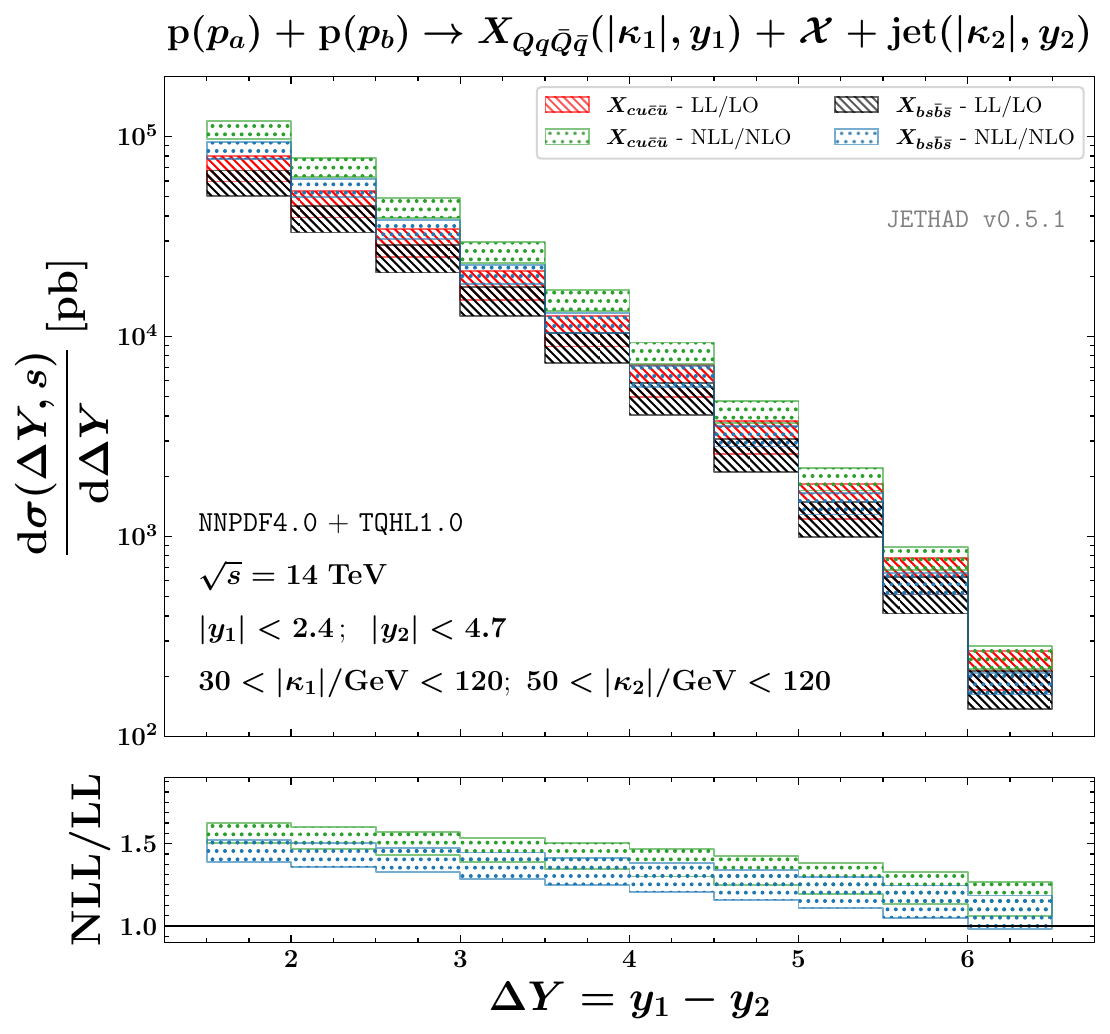}

\caption{Rapidity distribution for $\XQq + {\cal H}_{Q}$ (left) and $\XQq + {\rm jet}$ (right) production at $\sqrt{s} = 14$~TeV.}
\label{fig:rapidity_distribution}
\end{figure*}

\begin{figure*}[!t]
\centering

   \includegraphics[scale=0.41,clip]{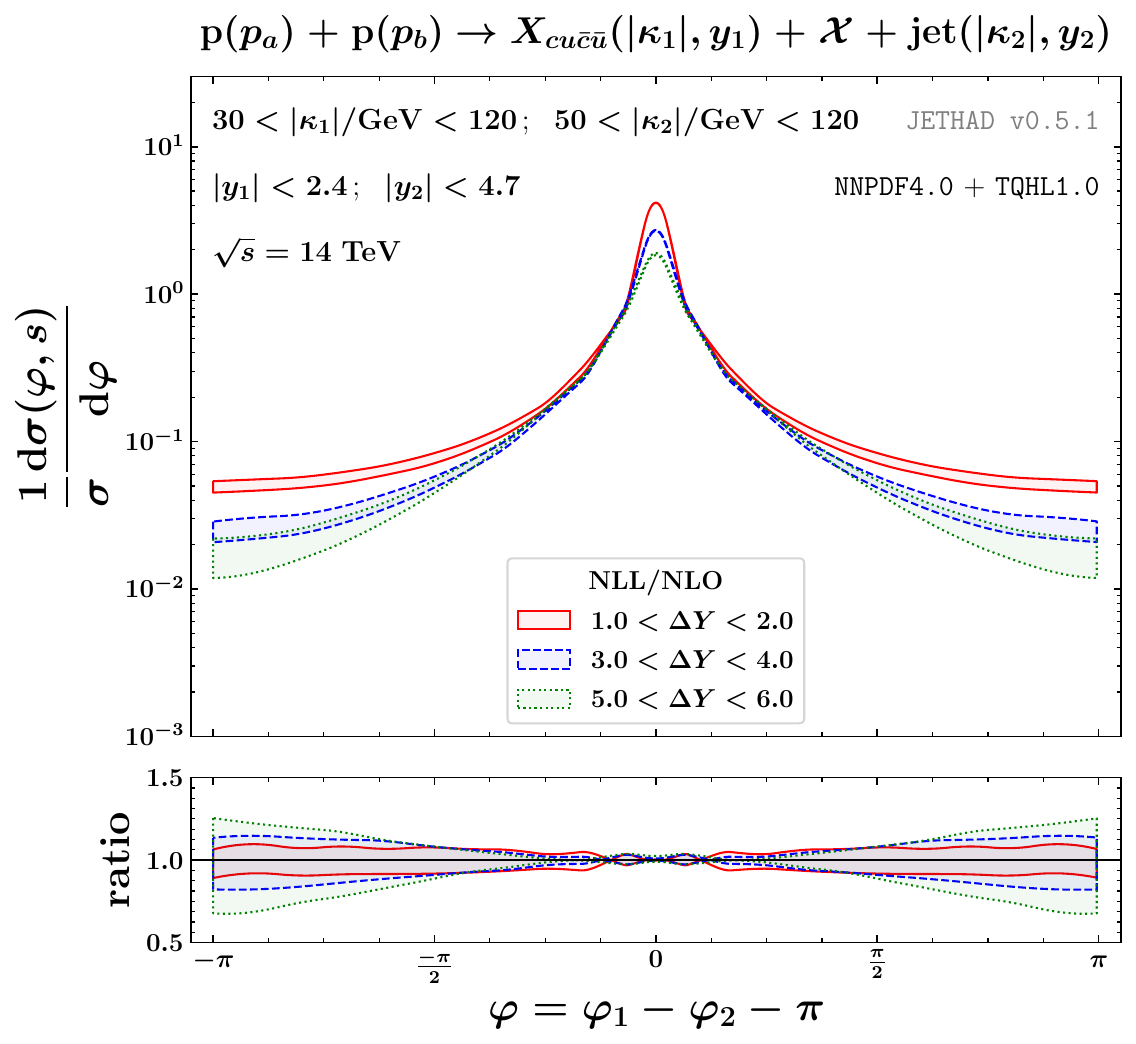}
   \hspace{0.10cm}
   \includegraphics[scale=0.41,clip]{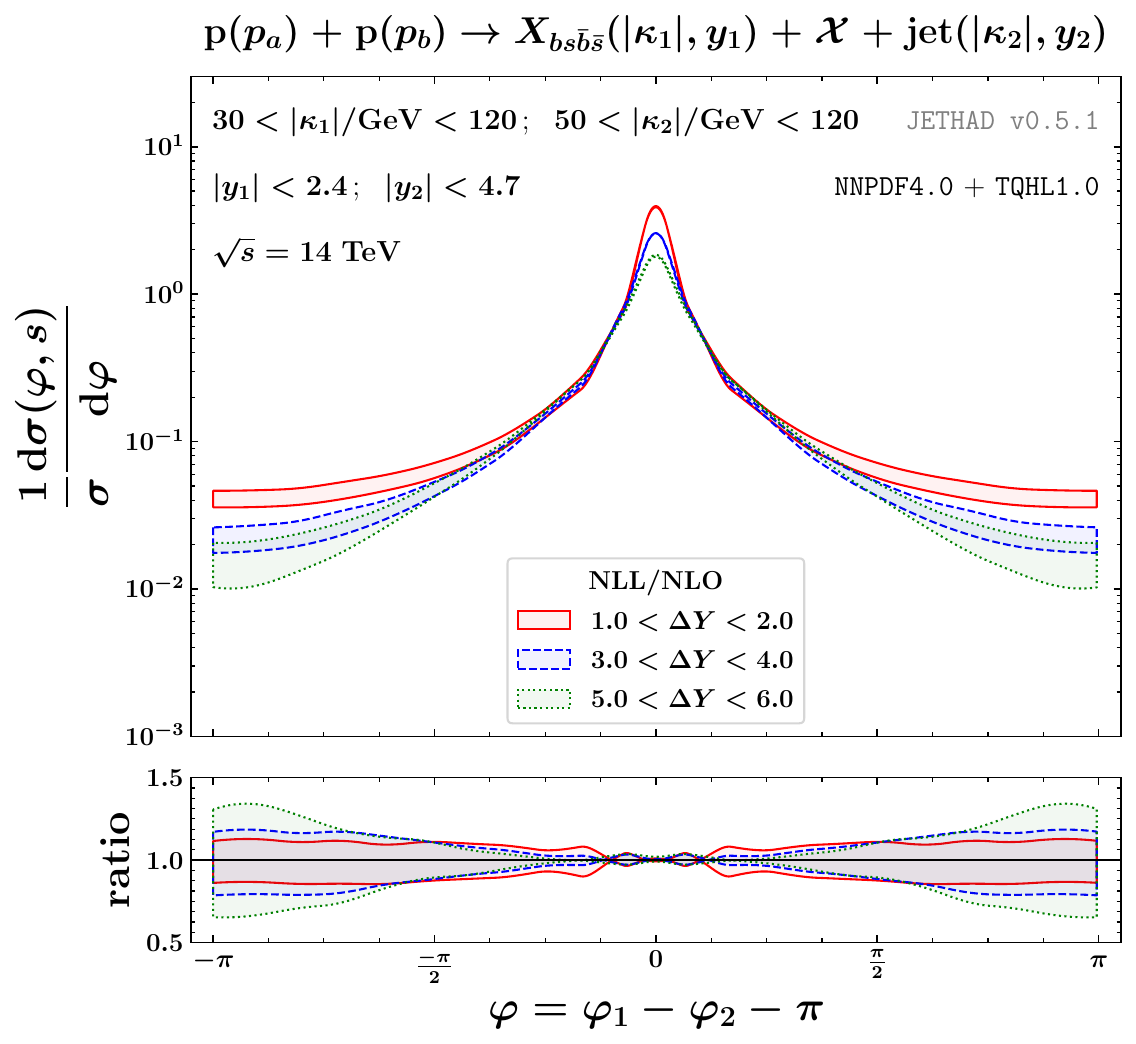}

\caption{NLL/NLO azimuthal distribution for $\Xcu + {\rm jet}$ (left) and $\Xbs + {\rm jet}$ (right) production at $\sqrt{s} = 14$~TeV.}
\label{fig:azimuthal_distribution}
\end{figure*}

We first investigate the $\DY$-differential distribution, obtained by integrating the r.h.s. of Eq.~\eqref{dsigma_Fourier} over the azimuthal-angle difference, $\varphi$. It corresponds to the first azimuthal coefficient, $C_0^{\rm int}$, obtained by integrating $C_0$ over a definite range of $|\bm{\kappa}_1|$ and $|\bm{\kappa}_2|$.
To be consistent with the fragmentation approximation~\cite{Kolodziej:1995nv,Artoisenet:2007xi}, we let tetraquark transverse momenta be in the range $30 < |\bm{\kappa}_1|/{\rm GeV} < 120$.
Then, to realize an asymmetric kinematic configuration, suited to better disentangle high-energy dynamics from fixed-order background~\cite{Ducloue:2013bva,Celiberto:2015yba,Celiberto:2020wpk}, ${\cal H}_{c,b}$ hadron or jet transverse momenta lie in the $50 < |\bm{\kappa}_2|/{\rm GeV} < 120$ window.
Our choice for $\bm{\kappa}_{1,2}$ cuts is compatible with a VFNS treatment, whose validity holds when energy scales are well above thresholds for DGLAP evolution dictated by heavy-quark masses.
To propose realistic configurations that can be easily compared with future HL-LHC data, we consider $\DY$-bins with a fixed length of 0.5.
The falloff with $\DY$ is the net effect of two competing behaviors: BFKL partonic cross sections grow with $\DY$ and then with energy, while their convolution with collinear PDFs and FFs in the impact factors strongly suppresses that growth.
Ancillary panels below main plots in Fig.\tref{fig:rapidity_distribution} highlight the stabilizing power of our tetraquark FFs. Here, NLL/NLO bands almost everywhere nested inside LL/LO ones, with NLL terms correcting pure LL results by a maximum factor of 10\% ($\XQq + {\cal H}_{Q}$, left) or 50\% ($\XQq + {\rm jet}$, right). This difference is expected, since in jet emissions the further stabilizer, namely $\cal{H}_{Q}$ FFs, is missing.
At variance with previous high-energy studies on heavy-flavor~\cite{Celiberto:2021dzy,Celiberto:2021fdp,Celiberto:2022dyf}, the NLL/LL ratio decreases and tends to one as $\DY$ grows, namely in the most BFKL-sensitive kinematic region.
This is further and stronger signal that heavy-light tetraquark production from fragmentation is a very stable channel to access high-energy QCD.

We then consider the azimuthal distribution, \emph{e.g.}, the normalized cross section differential in $\varphi$ and $\DY$,
\begin{eqnarray}
 \label{azimuthal_distribution}
 \frac{1}{\sigma} \frac{\drv \sigma}{\drv \varphi \, \drv \DY} = \frac{1}{2 \pi} + \frac{1}{\pi} \sum_{n=1}^\infty
 \langle \cos(n \varphi) \rangle \, \cos (n \varphi)\;,
\end{eqnarray}
the mean values $\langle \cos(n \varphi) \rangle$ being given as azimuthal-cor\-re\-lat\-ion moments, $\langle \cos(n \varphi) \rangle \equiv C_n^{\rm int}/C_0^{\rm int}$, with the $C_n$ integrated over the previously given $|\bm{\kappa}_{1,2}|$ ranges.
Since this distribution encodes signals from all azimuthal modes, it represents one of the fairest observables leading to the emergence the core high-energy dynamics.
For the sake of brevity, we present in Fig.\tref{fig:azimuthal_distribution} results for the NLL/NLO azimuthal distribution just in the $\XQq + {\rm jet}$ channel, which maximizes the $\DY$-range coverable.
We allow $\DY$ to span from one to six units and we select three bins of unit length.
As a general trend, our distributions are peaked around $\varphi = 0$, which kinematically corresponds to (almost) back-to-back emissions. The largest peak is at the lowest rapidity bin, $1 < \DY < 2$. As $\DY$ increases, the peak height decreases and the distribution width enlarges.
This is a fair indication that high-energy dynamics becomes more and more manifest when we approach the large-$\DY$ regime. Here, as expected, the weight of secondary gluons encoded in the ${\cal X}$ system~(Eq.\eref{process}) accounted for by BFKL grows. Thus, the number of back-to-back events diminishes and likewise the azimuthal correlation between the two outgoing objects.
As a bonus, we observe that computed $\varphi$-distributions are always positive, as expected.
Here, the stabilizing power rising from tetraquark fragmentation overpowers the oscillating instabilities typically affecting two large-$|\varphi|$ tails of the distributions (see~\cite{Celiberto:2022dyf,Celiberto:2022zdg,Celiberto:2022keu}) and connected to \emph{threshold} logarithmic contaminations~\cite{Celiberto:2022kxx}.
These Sudakov double logarithms, genuinely not accounted for by BFKL, may still survive and be responsible of the increasing uncertainty with $|\varphi|$ visible in ancillary panels of Fig.\tref{fig:azimuthal_distribution}.

\subsection{Transverse-momentum distributions}
\label{ssec:pT}

\begin{figure*}[!t]
\centering

   \includegraphics[scale=0.41,clip]{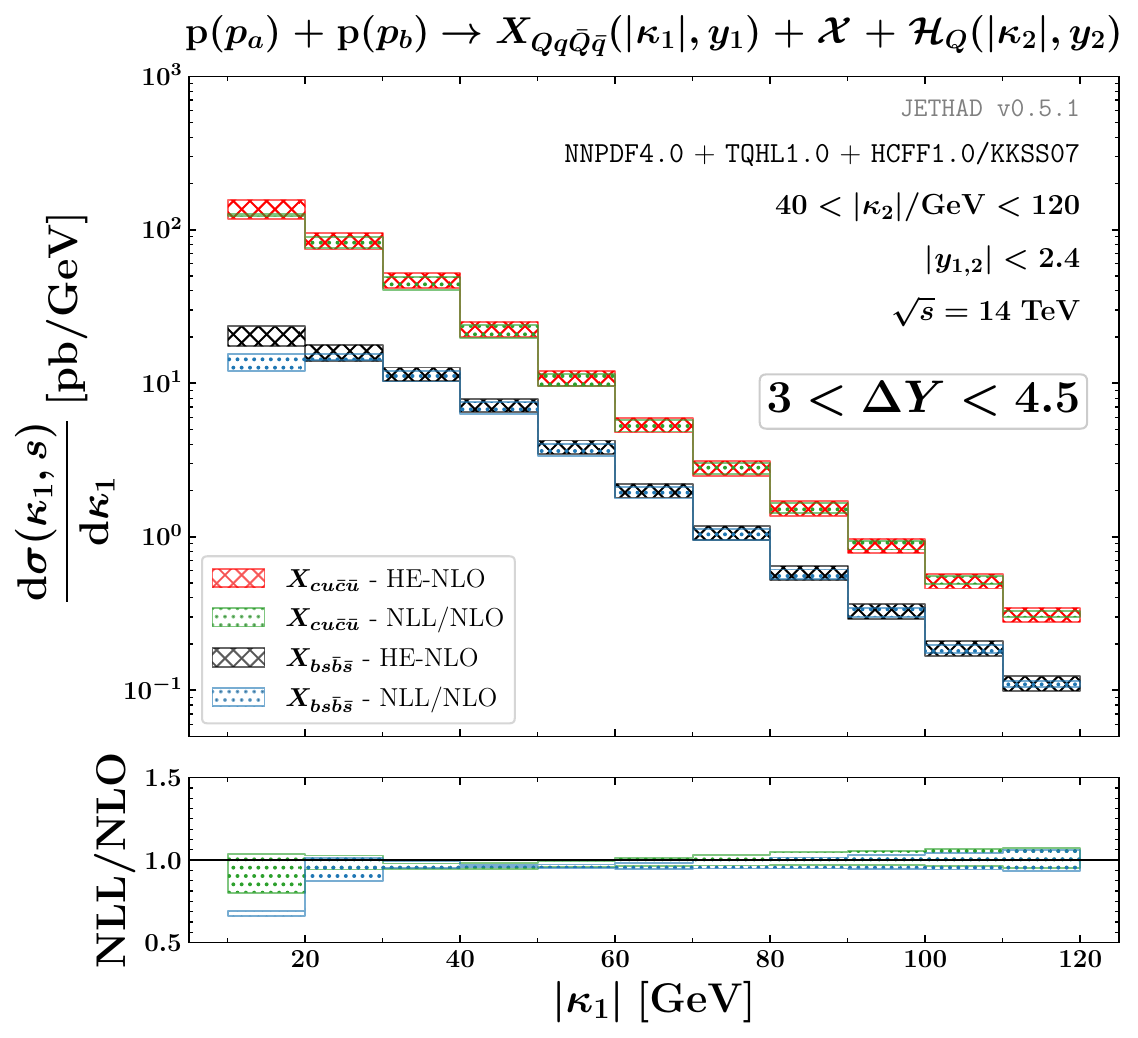}
   \hspace{0.10cm}
   \includegraphics[scale=0.41,clip]{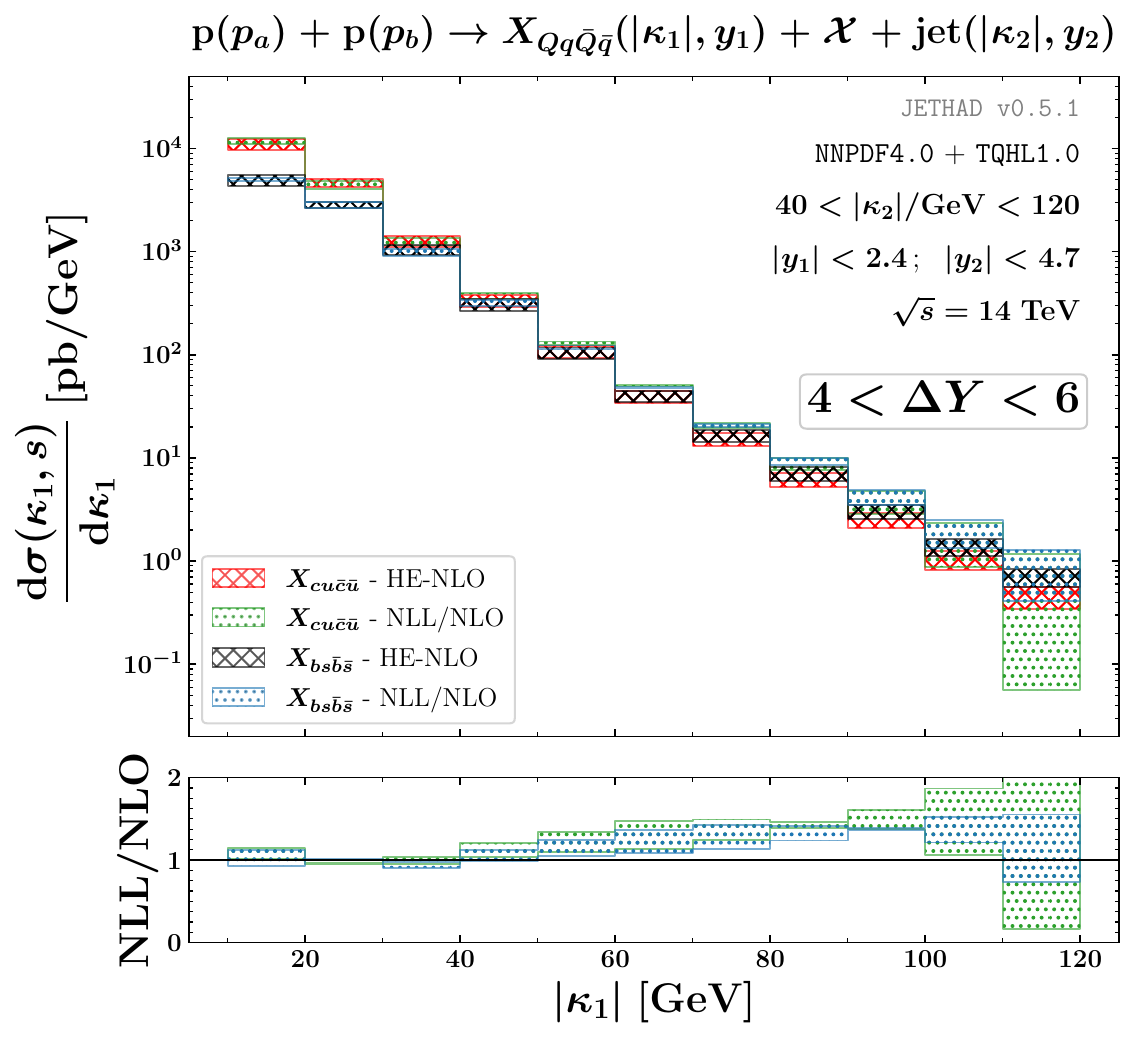}

   \includegraphics[scale=0.41,clip]{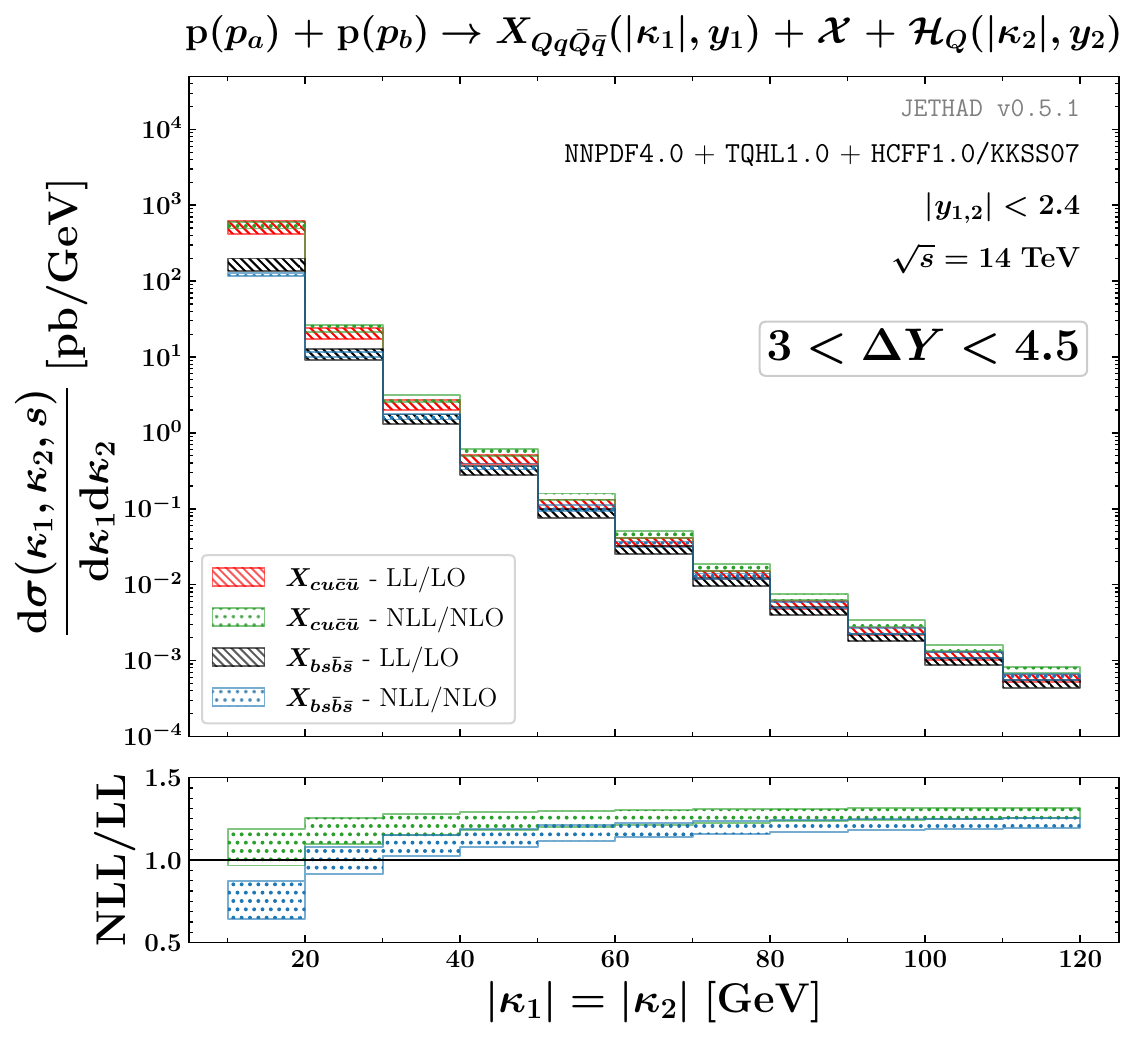}
   \hspace{0.10cm}
   \includegraphics[scale=0.41,clip]{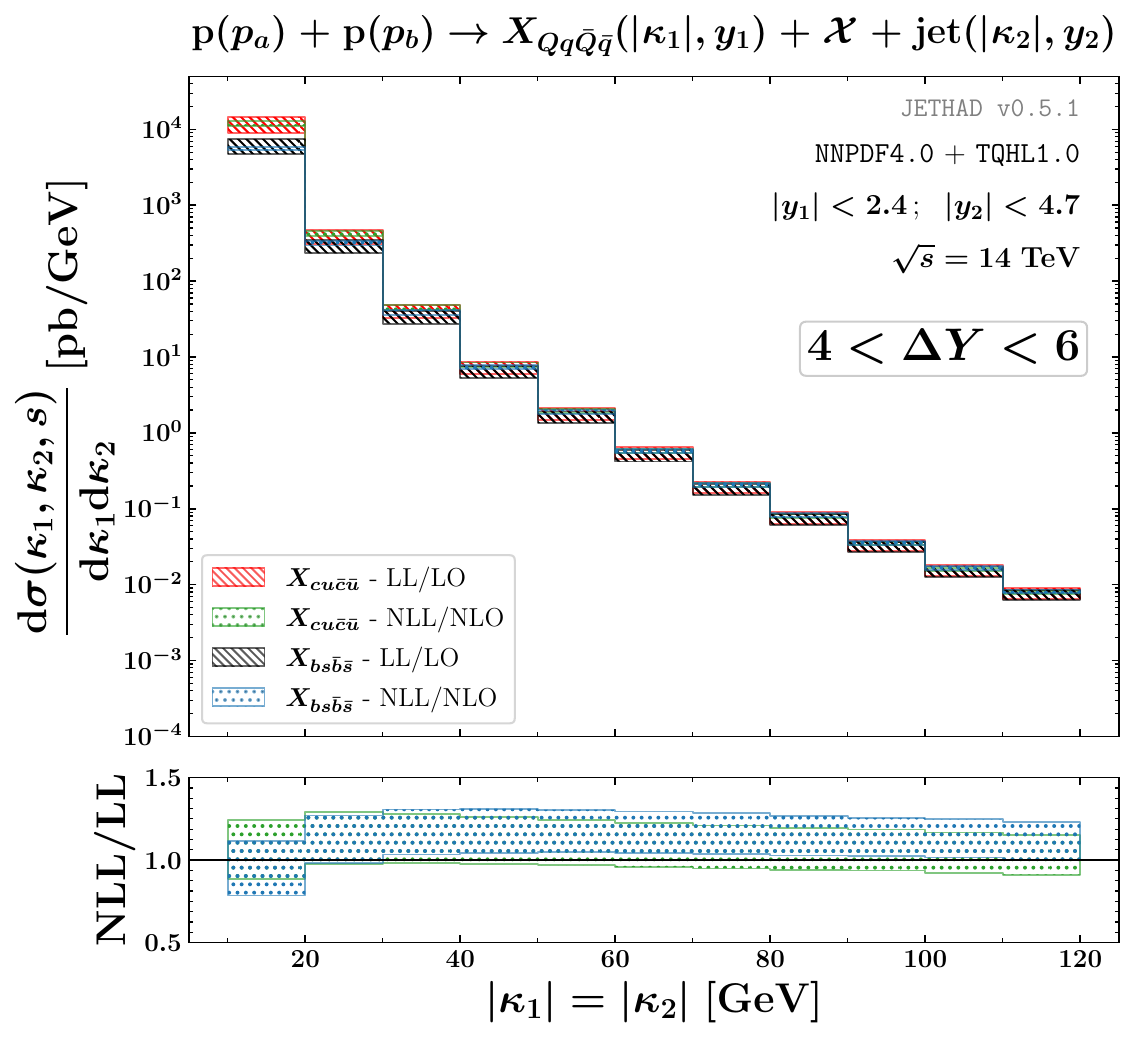}

\caption{Transverse-momentum distributions for $\XQq + {\cal H}_{Q}$ (left) and $\XQq + {\rm jet}$ (right) production at $\sqrt{s} = 14$~TeV.}
\label{fig:pT_distributions}
\end{figure*}

The first distribution under investigation is the $|\bm{\kappa}_1|$-differential cross section, obtained by integrating $C_0$ over $|\bm{\kappa}_2|$ from 40 to 120~GeV.
We integrate $\DY$ in a process-dependent forward bin, namely $3 < \DY < 4.5$ ($\XQq + {\cal H}_{Q}$ channel) or $4 < \DY < 6$ ($\XQq + {\rm jet}$ channel).
This observable gives us a common ground whereby unveiling connections between our formalism and other approaches. Indeed, letting $|\bm{\kappa}_1|$ range from 10 to 120~GeV allows one to scan a wide kinematic sector where distinct resummation mechanisms progressively come into play. Recent results on high-energy Higgs~\cite{Celiberto:2020tmb} and heavy-jet~\cite{Bolognino:2021mrc} emissions at the LHC have shown how the BFKL resummation works well in the moderate transverse-momentum subregion, when $|\bm{\kappa}_1| \sim |\bm{\kappa}_2|$, while large $|\bm{\kappa}_1|$ and threshold logarithms are expected to rise in the soft ($|\bm{\kappa}_1| \ll |\bm{\kappa}_2^{\rm min}|$) and hard ($|\bm{\kappa}_1| \sim |\bm{\kappa}_2^{\rm max}|$) subregions, respectively.
In this study we investigate $|\bm{\kappa}_1|$-distributions to test the stability of our factorization and possibly trace the path toward futures studies aimed at shedding light on the interplay among different resummations.
In upper plots of Fig.\tref{fig:pT_distributions} we compare NLL/NLO $|\bm{\kappa}_1|$-distributions in the double-hadron (left) and hadron-plus-jet channel (right) with the corresponding HE-NLO limit~(Eq.\eref{Cn_HENLO_MSbar}).
The overall pattern is a net falloff with $|\bm{\kappa}_1|$.
Results are very stable under energy-scale variations, with error bands not exceeding 30\% width except for the first and the last two bins.
From ancillary panels showing the ratio between the two calculations, labeled as NLL/NLO, we note that, in the double-hadron case, the resummed distribution is smaller than the fixed-order one in the first bin, then it reaches the same order and it slightly grows with $|\bm{\kappa}_1|$. 
\emph{Vice versa}, in the hadron-plus-jet case, the resummation leads to a visible increase with $|\bm{\kappa}_1|$, up to 50\%.
Then, larger uncertainties in the last two bins clearly indicate that BFKL starts losing stability due to the aforementioned threshold contaminations. These uncertainties are less pronounced when a $\Xbs$ tetraquark is detected. This is in line with recent findings corroborating the statement that VFNS FFs depicting bottom-flavored hadrons carry a stronger stabilizing power than charm-flavored ones~\cite{Celiberto:2021fdp}.
On the other side, the first bin could be affected by instabilities due to the fact that energy scales are rapidly approaching thresholds for DGLAP evolution given by heavy-quark masses. This could explain the turn-down of the $\Xbs$ NLL cross section in the left plot, which could suffer from instabilities due to values of $\mu_F$ close to the bottom mass.

The last observable investigated is a distribution doubly differential both in $|\bm{\kappa}_1|$ and $|\bm{\kappa}_2|$.
Here, as the distance between the two transverse momenta increases, other kinematic sectors, contiguous to the BFKL one, are accessed (see recent analyses on ${\cal H}_b$ hadrons~\cite{Celiberto:2021fdp} and $\Xi$ baryons~\cite{Celiberto:2022kxx}).
A joint resummation of transverse-momentum logarithms for two-particle distributions was afforded for the first time in the context of Higgs-plus-jet hadroproduction~\cite{Monni:2019yyr} \emph{via} the {\tt RadISH} momentum-space method~\cite{Bizon:2017rah}.
In this study we pick a complementary configuration, namely we set $|\bm{\kappa}_1|$ = $|\bm{\kappa}_2|$ and let them span from 10 to 120~GeV. Rapidity bins are the same as before.
This choice allows us to deeply focus on a strict BFKL regime, hunting for a precise determination of the impact of NLL corrections.
Lower plots of Fig.\tref{fig:pT_distributions} 
are for NLL/NLO distributions in the double hadron (left) and hadron-plus-jet channel (right), compared with their LL/LO limits.
The NLL to LL ratio is magnified in ancillary plots.
Apart from the first bin, where $\Xbs$ cross sections suffer from the aforementioned threshold-proximity instabilities, NLL corrections in the left plot moderately increase with the transverse momentum, up to reach a plateau at around +30\%.
Conversely, NLL terms in the right plots corrects LL results by +30\% in the whole spectrum.
This dichotomy originates from the fact that NLO hadron impact-factor corrections are positive due to large values of the $(gg)$ channel~\cite{Celiberto:2017ptm}, while NLO jet ones are always negative~\cite{Ducloue:2013bva,Caporale:2014gpa,Celiberto:2015yba,Celiberto:2022gji}.

\section{Future perspectives}
\label{sec:conclusions}

By relying upon the hybrid high-energy and collinear factorization, where the BFKL resummation of energy logarithms accompanies and enhances the standard collinear description, we provided a prime study on the inclusive hadroproduction of $\XQq$ tetraquark states at high energies, emitted in association with single-charmed hadrons or light jets.
We described the tetraquark production mechanism \emph{via} the fragmentation approximation, valid in the large transverse-momentum regime matter of our analysis.
To this extent, we created a first and novel tetraquark collinear NLO FF set, named {\tt TQHL1.0}. It was built by evolving \emph{\`a la} DGLAP a Suzuki-like model input for the heavy-quark function~\cite{Nejad:2021mmp}.
A key role is played by the interplay between the high-energy resummation and the fragmentation mechanism.
A striking evidence came out that the peculiar behavior of the gluon to tetraquark fragmentation channel acts as a fair \emph{stabilizer} of the hybrid factorization under NLL corrections and scale variations, thus permitting us to reach a remarkable level of accuracy in the description of our observables at the natural scales provided by kinematics. 
This \emph{natural stability}, emerged as a general property carried by the heavy-flavored species considered so far (single $c$- or $b$-particles~\cite{Celiberto:2021dzy,Celiberto:2021fdp,Celiberto:2022rfj,Celiberto:2022zdg}, quarkonia~\cite{Celiberto:2022dyf}, and charmed $B$ mesons~\cite{Celiberto:2022keu}), has now been observed also in the case of $\XQq$ states.

Future extensions of this work are needed to address heavy-flight tetraquark hadroproduction by other resummation mechanisms, as well as by different inputs for the initial-scale fragmentation and through exclusive reactions.
Novel, promising channels to be accessed \emph{via} the fragmentation approximation are fully-charmed tetraquark~\cite{Feng:2020riv,Feng:2020qee} and pentaquark~\cite{Cheung:2004ji} emissions.
Until recently, the $X(3872)$ was the only exotic state to be observed in prompt proton collisions. The discovery of fully-charmed structures~\cite{LHCb:2020bwg} and of the $T_{cc}$~\cite{LHCb:2021vvq,LHCb:2021auc} changed the situation.
The extension of our fragmentation approach to them should be straightforward.
It could also be employed to study the $Z_c(3900)$, which has not been observed promptly so far~\cite{Guo:2013ufa}.
Though further analyses are still needed both from the formal and phenomenological sides, we believe that a new high-energy QCD portal to shed light on the true nature of the exotic matter is now open and accessible at the forthcoming HL-LHC.

\section*{Data availability}

Data are publicly available from \url{https://github.com/FGCeliberto/Collinear_FFs/}.
We deliver the NLO FF sets in {\tt LHAPDF} format for our tetraquark states (see the {\tt TQHL10/} subfolder):
\begin{itemize}
    \item NLO, $\Xcu$\,: \,{\tt TQHL10\_Xcu\_nlo};
    \item NLO, $\Xcs$\,: \,{\tt TQHL10\_Xcs\_nlo};
    \item NLO, $\Xbu$\,: \,{\tt TQHL10\_Xbu\_nlo};
    \item NLO, $\Xbs$\,: \,{\tt TQHL10\_Xbs\_nlo}.
\end{itemize}

\section*{Acknowledgements}

We are grateful to colleagues of the \textbf{Quarkonia As Tools} and \textbf{EXOTICO} Workshops for the fruitful conversations and the warm atmosphere which inspired us to realize the present study.
We thank A.~D'Angelo, A.~Pilloni, I.~Schienbein, and S.~M.~Moosavi~Nejad for insightful discussions.
This work was supported by the Atracci\'on de Talento Grant n. 2022-T1/TIC-24176 of the Comunidad Aut\'onoma de Madrid, Spain, and by the INFN/QFT@COLLIDERS Project, Italy.

\bibliographystyle{elsarticle-num}

\bibliography{bibliography}

\end{document}